\documentclass[a4paper]{jpconf}
\usepackage{graphicx}
\begin{document}
\title{The {\sc Majorana} Project}

\newcommand{\nnprime}{n,n$^\prime \gamma$}
\newcommand{\ntwon}{n,2n$\gamma$}
\newcommand{\nthreen}{n,3n$\gamma$}
\newcommand{\nxn}{n,xn$\gamma$}
\newcommand{\nx}{n,x$\gamma$}

\newcommand{\natpb}{$^{\textrm{nat}}$Pb}
\newcommand{\natge}{$^{\textrm{nat}}$Ge}
\newcommand{\eigpb}{$^{208}$Pb}
\newcommand{\sevpb}{$^{207}$Pb}
\newcommand{\sixpb}{$^{206}$Pb}
\newcommand{\fivpb}{$^{205}$Pb}
\newcommand{\foupb}{$^{204}$Pb}
\newcommand{\nonubb}  {$0 \nu \beta \beta$}
\newcommand{\twonubb} {$2 \nu \beta \beta$}
\newcommand{\gam}{$\gamma$}
\def\nuc#1#2{${}^{#1}$#2}
\def\mee{$\langle m_{\beta\beta} \rangle$}
\def\mnu{$\langle m_{\nu} \rangle$}
\def\ml{$m_{lightest}$}
\def\gnu{$\langle g_{\nu,\chi}\rangle$}
\def\mmod{$\| \langle m_{\beta\beta} \rangle \|$}
\def\mb{$\langle m_{\beta} \rangle$}
\def\BBz{$0 \nu \beta \beta$}
\def\BBm{$\beta\beta(0\nu,\chi)$}
\def\BBt{$2 \nu \beta \beta$}
\def\nonubb{$0 \nu \beta \beta$}
\def\twonubb{$2 \nu \beta \beta$}
\def\BB{$\beta\beta$}
\def\Mz{$M_{0\nu}$}
\def\Mt{$M_{2\nu}$}
\def\MzG{$M^{GT}_{0\nu}$}           
\def\MzF{$M^{F}_{0\nu}$}                
\def\MtG{$M^{GT}_{2\nu}$}           
\def\MtF{$M^{F}_{2\nu}$}                
\def\Tz{$T^{0\nu}_{1/2}$}
\def\Tt{$T^{2\nu}_{1/2}$}
\def\Tc{$T^{0\nu\,\chi}_{1/2}$}
\def\Rz{$\Gamma_{0\nu}$}            
\def\Rt{$\Gamma_{2\nu}$}            
\def\ms{$\delta m_{\rm sol}^{2}$}
\def\ma{$\delta m_{\rm atm}^{2}$}
\def\ts{$\theta_{\rm sol}$}
\def\ta{$\theta_{\rm atm}$}
\def\tot{$\theta_{13}$}
\def\gpp{$g_{pp}$}                  
\def\qval{$Q_{\beta\beta}$}                 
\def\MJ{{\sc Majorana}}             
\def\DEM{{\sc Demonstrator}}             
\def\be{\begin{equation}}
\def\ee{\end{equation}}
\def\cpRty{counts/ROI/t-y}
\def\onecpRty{1~count/ROI/t-y}
\def\fourcpRty{4~counts/ROI/t-y}
\def\ppc{P-PC}                          
\def\nsc{N-SC}                          

\newcommand{\alberta}{Centre for Particle Physics, University of Alberta, Edmonton, AB, Canada}
\newcommand{\blhill}{Department of Physics, Black Hills State University, Spearfish, SD, USA}
\newcommand{\ITEP}{Institute for Theoretical and Experimental Physics, Moscow, Russia}
\newcommand{\JINR}{Joint Institute for Nuclear Research, Dubna, Russia}
\newcommand{\lbnl}{Lawrence Berkeley National Laboratory, Berkeley, CA, USA}
\newcommand{\lanl}{Los Alamos National Laboratory, Los Alamos, NM, USA}
\newcommand{\queens}{Department of Physics, Queen's University, 
Kingston, ON, Canada}
\newcommand{\uw}{Center for Experimental Nuclear Physics and Astrophysics, 
and Department of Physics, University of Washington, Seattle, WA, USA}
\newcommand{\uchic}{Department of Physics, University of Chicago, Chicago, IL, USA}
\newcommand{\unc}{Department of Physics, University of North Carolina, Chapel Hill, NC, USA}
\newcommand{\ucne}{Department of Nuclear Engineering, University of California, Berkeley, CA, USA}
\newcommand{\ucph}{Department of Physics, University of California, Berkeley, CA, USA}
\newcommand{\duke}{Department of Physics, Duke University, Durham, NC, USA}
\newcommand{\ncsu}{Department of Physics, North Carolina State University, Raleigh, NC, USA}
\newcommand{\ornl}{Oak Ridge National Laboratory, Oak Ridge, TN, USA}
\newcommand{\ou}{Research Center for Nuclear Physics and Department of Physics, Osaka University, Ibaraki, Osaka, Japan}
\newcommand{\pnnl}{Pacific Northwest National Laboratory, Richland, WA, USA}
\newcommand{\usc}{Department of Physics and Astronomy, University of South Carolina, Columbia, SC, USA}
\newcommand{\usd}{Department of Earth Science and Physics, University of South Dakota, Vermillion, SD, USA}
\newcommand{\ut}{Department of Physics and Astronomy, University of Tennessee, Knoxville, TN, USA}
\newcommand{\tunl}{Triangle Universities Nuclear Laboratory, Durham, NC, USA}

\author{C.E.~Aalseth$^{10}$,
M.~Amman$^{6}$,
J.F.~Amsbaugh$^{20}$,
F.T.~Avignone~III$^{8,17}$,
H.O.~Back$^{7,11}$,
A.S.~Barabash$^{3}$,
P.S.~Barbeau$^{15}$,
J.R.~Beene$^{17}$,
M.~Bergevin$^{6}$,
F.E.~Bertrand$^{8}$,
M.~Boswell$^{5}$, 
V.~Brudanin$^{4}$,
W.~Bugg$^{19}$,
T.H.~Burritt$^{20}$,
M.~Busch$^{2,11}$,	
G.~Capps$^{8}$,
Y-D.~Chan$^{6}$,
J.I.~Collar$^{15}$,
R.J.~Cooper$^{8}$,
R.~Creswick$^{17}$,
J.A.~Detwiler$^{6}$,
P.J.~Doe$^{20}$,
Yu.~Efremenko$^{19}$,
V.~Egorov$^{4}$,
H.~Ejiri$^{9}$,
S.R.~Elliott$^{5}$,
J.~Ely$^{10}$,
J.~Esterline$^{2,11}$,
H.~Farach$^{17}$,
J.E.~Fast$^{10}$,
N.~Fields$^{15}$,
P.~Finnerty$^{11,16}$,
B.~Fujikawa$^{6}$,
E.~Fuller$^{10}$,
V.M.~Gehman$^{5}$,
G.K.~Giovanetti$^{11,16}$,
V.E.~Guiseppe$^{5}$,
K.~Gusey$^{4}$,
A.L.~Hallin$^{12}$,
R.~Hazama$^{9}$,
R.~Henning$^{11,16}$,
A.~Hime$^{5}$,
E.W.~Hoppe$^{10}$,
T.W.~Hossbach$^{10}$,
M.A.~Howe$^{11,16}$,
R.A.~Johnson$^{20}$,
K.J.~Keeter$^{1}$,
M.~Keillor$^{10}$,
C.~Keller$^{18}$,
J.D.~Kephart$^{7,10,11}$,
M.F.~Kidd$^{2,11}$,
O.~Kochetov$^{4}$,
S.I.~Konovalov$^{3}$,
R.T.~Kouzes$^{10}$,
L.~Leviner$^{7,11}$,
J.C.~Loach$^{6}$,
P.N.~Luke$^{6}$,
S.~MacMullin$^{11,16}$,
M.G.~Marino$^{20}$,
R.D.~Martin$^{6}$,
D.-M.~Mei$^{18}$,
H.S.~Miley$^{10}$,
M.L.~Miller$^{20}$,
L.~Mizouni$^{10,17}$,
A.~Montoya$^{5}$,
A.W.~Myers$^{10}$,
M.~Nomachi$^{9}$,
J.L.~Orrell$^{10}$,
D.G.~Phillips II$^{11,16}$,
A.W.P.~Poon$^{6}$,
G.~Prior$^{6}$,
J. ~Qian$^{6}$,
D.C.~Radford$^{8}$,
K.~Rielage$^{5}$,
R.G.H.~Robertson$^{20}$,
L.~Rodriguez$^{5}$,
K.P.~Rykaczewski$^{8}$,
A.G.~Schubert$^{20}$,
T.~Shima$^{9}$,
M.~Shirchenko$^{4}$,
D.~Steele$^{5}$,
J.~Strain$^{11,16}$,
G.~Swift$^{2,11}$,
K.~Thomas$^{18}$,	
R.~Thompson$^{10}$,
V.~Timkin$^{4}$,
W.~Tornow$^{2,11}$,
T.D.~Van Wechel$^{20}$,
I.~Vanyushin$^{3}$,
K.~Vetter$^{6,13}$,
R.~Warner$^{10}$,
J.F.~Wilkerson$^{11,16}$,
J.M.~Wouters$^{5}$,
E.~Yakushev$^{4}$,
A.R.~Young$^{7,11}$,
C.-H.~Yu$^{8}$,
V.~Yumatov$^{3}$,
C.~Zhang$^{18}$,					
S.~Zimmerman$^{6}$
}
			
\address{$^{1}$ \blhill}
\address{$^{2}$ \duke}
\address{$^{3}$ \ITEP}
\address{$^{4}$ \JINR}
\address{$^{5}$ \lanl}
\address{$^{6}$ \lbnl}
\address{$^{7}$ \ncsu}
\address{$^{8}$ \ornl}
\address{$^{9}$ \ou}
\address{$^{10}$ \pnnl}
\address{$^{11}$ \tunl}
\address{$^{12}$ \alberta}
\address{$^{13}$ \ucne}
\address{$^{14}$ \ucph}
\address{$^{15}$ \uchic}
\address{$^{16}$ \unc}
\address{$^{17}$ \usc}
\address{$^{18}$ \usd}
\address{$^{19}$ \ut}
\address{$^{20}$ \uw}

\ead{elliotts@lanl.gov}

\begin{abstract}
The {\sc Majorana} Project, a neutrinoless double-beta decay experiment is described with an emphasis on the choice of Ge-detector configuration.
\end{abstract}

Experimental evidence of \BBz\
would establish the Majorana nature of neutrinos. The science of \BB\ and the large number of experimental and theoretical efforts in the field are described in detail, not only in this volume, but also in
many recent reviews~\cite{ell02,ell04,bar04,eji05,avi05,avi08}.

The objective of the first experimental phase of \MJ\
is to build a 60-kg module of high-purity Ge, of which 30 kg will be enriched to 86\% in $^{76}$Ge,
to search \BBz.  This module is referred to by the collaboration as the \DEM. The physics goals 
for this first phase are to: probe the neutrino mass region above 100~meV;
demonstrate that backgrounds at or below
1~count/tonne-year in the \BBz-decay region of interest 
can be achieved that would justify scaling up to a 1~tonne or larger
mass detector; and definitively test the recent claim~\cite{kla06}  
of an observation of \BBz\ decay in $^{76}$Ge in the mass region around
400~meV.

The \DEM\ will
consist of $^{76}$Ge  detectors, deployed in multi-crystal modules,
located deep underground within a low-background shielding environment. The technique 
will be augmented with recent improvements in
signal processing and detector design, and advances in controlling
intrinsic and external backgrounds.  
Recently, R\&D by the \MJ\ and GERDA~\cite{sch05} collaborations have shown the utility of p-type, point-contact (\ppc) Ge detectors for \BB. \MJ\ has been described elsewhere~\cite{ell08,gui08,hen09}, so this short note will summarize the efforts developing these detectors.

The efficient commercial production of Ge detectors depends strongly
on the yields for producing high-purity single crystals of Ge and for
producing p-n junction diodes from these crystals.  Single-crystal
boules naturally favor right circular cylindrical geometries, making
such an obvious choice to make efficient use the precious
enriched Ge material. 

Previous-generation \BB\ decay searches in $^{76}$Ge favored
detectors made of p-type crystals that
can be grown more efficiently to larger dimensions than n-type
crystals.  In addition, charge-trapping effects are
reduced, and generally slightly better energy resolution is obtained.
P-type detectors are typically fabricated with a p+ B-implanted
contact on an inner bore hole and an n+ Li-drifted contact on the
outside electrode to obtain efficient and full depletion from the
outside.  While the inside B contact is very thin (typically
$<$1~$\mu$m) the Li-drifted contact can be as thick as 1~mm due to the high mobility of Li
in Ge. The thickness of the outside-Li contact provides an advantage
since this dead layer absorbs $\alpha$ radiation
from surface contamination that might generate background.

While the pulse-shape obtained at the central contact in coaxial
detectors can provide radial separation of multiple interactions in
any implementation, pulse-shapes obtained in segmented detectors
provide improved sensitivity in the radial separation and, more
importantly, in complementary directions as well. A high degree of
segmentation, such as a 6x6-fold segmentation, enables the full
reconstruction of $\gamma$-ray interactions within the detectors.
Even without absolute event vertex
reconstruction, events with multiple energy depositions can be
identified and rejected.  Preliminary results from GRETINA~\cite{Vet00} and other
highly-segmented HPGe arrays indicate that a minimum separation of
4~mm will be achievable.  However, these advanced capabilities come at
the price of a proportionally larger number of small parts such as
cables and FETs, with their associated additional sources of
background.

An alternative cylindrical detector design has been
developed in which the bore hole is removed and replaced with a
point-contact in the center of the passivated detector
face~\cite{Luk89}.  The changes in the electrode structure result in a
drop in capacitance to $\sim$1~pF, reducing the electronic noise
component and enabling sub-keV energy thresholds.  This configuration
also has lower electric fields throughout the bulk of the crystal and
a weighting potential that is sharply peaked near the point contact,
resulting in an increased range of drift times and a distinct electric
signal marking the arrival of the charge cloud at the central
electrode. These features result in an improved pulse-shape analysis (PSA) over semi-coax detectors that rivals the performance of segmented detectors.  Hence, p-type versions of this configuration are our favored choice.  Instrumented with modern FETs, \ppc\ detectors have recently
been demonstrated by \MJ\ collaborators to provide very low noise and
energy threshold, as well as excellent pulse-shape capabilities for
distinguishing multiple interactions from single
interactions~\cite{Barb07}.

In our R\&D efforts, we have studied a large number of detector designs. Many of these
detectors were custom made to our specifications either by vendors or in-house to explore
various performance aspects. We studied segmented detectors with a commercial CLOVER
detector, a custom ORTEC-fabricated~\cite{ORTEC} enriched-Ge segmented detector (SEGA), a Canberra-fabricated highly segmented GRETINA prototype, and a LBNL-fabricated segmented \ppc. We have studied passivation issues with a LBNL-fabricated \ppc, and two detectors with various ditch geometries fabricated at PHDs~\cite{PHDs}. We also modified an n-type detector for surface $\alpha$-emitter studies. Recently, a novel variation on the \ppc\ detector design has been
proposed~\cite{Rad08} which makes use of a coaxial hole similar in
dimensions to that of a coaxial HPGe detector.  In this case, however,
the hole is part of the outer Li contact.
In these coaxial-style \ppc\ detectors, the hole serves to reduce the
depletion voltage of a large detector by about a factor of two,
allowing the use of much longer crystals, and crystals with higher
concentrations of electrical impurities.

Broad Energy Ge (BEGe) detectors~\cite{BEGe} are commercially
available. These detectors share many of
the features of \ppc\ detectors, and generally meet the requirements
for the \DEM. They have been produced with diameters up to
90~mm, but have a shorter maximum length of about 30~mm. We have studied a variety of
sizes of these detectors. We have chosen a BEGe design that is 70-mm diameter, 30-mm high to be the primary detector configuration for the \DEM\ due to its cost effectiveness, PSA-background rejection capabilities, and simplicity of the contacts. GERDA has also recently reported a BEGe study~\cite{dusa09}. We have purchased an initial batch of 18 of these detectors.  They have an average leakage current of 3 pA, capacitance 1.3 pF, and a resolution of 1.8 keV at 1332 keV.

\section*{References}
\bibliographystyle{iopart-num}
\bibliography{DoubleBetaDecay.bbl}

\providecommand{\newblock}{}
\begin{thebibliography}{10}
\expandafter\ifx\csname url\endcsname\relax
  \def\url#1{{\tt #1}}\fi
\expandafter\ifx\csname urlprefix\endcsname\relax\def\urlprefix{URL }\fi
\providecommand{\eprint}[2][]{\url{#2}}

\bibitem{ell02}
Elliott S~R and Vogel P 2002 {\em Ann.\ Rev.\ Nucl.\ Part.\ Sci.\/} {\bf 52}
  115

\bibitem{ell04}
Elliott S~R and Engel J 2004 {\em J.\ Phys.\ G:\ Nucl.\ Part.\ Phys.\/} {\bf
  30} R 183

\bibitem{bar04}
Barabash A~S, Hubert F, Huber P and Umatov V~I 2004 {\em Phys. At. Nucl.\/}
  {\bf 67} 438

\bibitem{eji05}
Ejiri H 2005 {\em J. Phys. Soc. Jap.\/} {\bf 74} 2101

\bibitem{avi05}
Avignone F~T~{\protect III}, King G~S~{\protect III} and Zdesenko Y 2005 {\em
  New Journal of Physics\/} {\bf 7} 6

\bibitem{avi08}
Avignone F~T~{\protect III}, Elliott S and Engel J 2008 {\em Rev. Mod. Phys.\/}
  {\bf 80} 481 (\textit{Preprint} \eprint{arXiv:0708.1033})

\bibitem{kla06}
Klapdor-Kleingrothaus H~V and Krivosheina I~V 2006 {\em Mod. Phys. Lett. A\/}
  {\bf 21} 1547

\bibitem{sch05}
Sch{\protect\"{o}}nert S {\em et~al.\/} 2005 {\em Nucl. Phys. Proc. Suppl.\/}
  {\bf 145} 242

\bibitem{ell08}
Elliott S {\em et~al.\/} 2008  (\textit{Preprint} \eprint{arXiv:0807.1741})

\bibitem{gui08}
Guiseppe V {\em et~al.\/} 2008 (\textit{Preprint} \eprint{arXiv:0811.2446})

\bibitem{hen09}
Henning R {\em et~al.\/} 2009 (\textit{Preprint} \eprint{arXiv:0907.1581})

\bibitem{Vet00}
Vetter K {\em et~al.\/} 2000 {\em Nuclear Instruments and Methods in Physics
  Research A\/} {\bf 452} 105--114

\bibitem{Luk89}
Luke P, Goulding F, Madden N and Pehl R 1989 {\em IEEE trans. Nucl. Sci.\/}
  {\bf 36} 926

\bibitem{Barb07}
Barbeau P, Collar J and Tench O 2007 {\em JCAP\/} {\bf 09} 009

\bibitem{ORTEC}
 2009 {O}RTEC, 801 South Illinois Avenue Oak Ridge, TN 37830, USA

\bibitem{PHDs}
 2009 {P}HDs Co. 777 Emory Valley Road, Suite B, Oak Ridge, TN 37830, USA

\bibitem{Rad08}
Radford D private communication

\bibitem{BEGe}
 2009 {B}road Energy Ge Detectors, Canberra Industries, Inc., 800 Research
  Pkwy, Meriden, CT 06450, USA

\bibitem{dusa09}
Budj{\protect\'{a}\protect\v{s}} D, Heider M~B, Chkvorets O, Khanbekov N and
  Sch{\protect\"{o}}nert S 2009 (\textit{Preprint} \eprint{arXiv:0909.4044})

\end{thebibliography}

\end{document}